\documentclass[pra,aps,twocolumn]{revtex4-2}
\usepackage{graphicx,amsmath,color,soul}

\begin{document}

\title{Effects of ultrafast free-carrier dynamics on frequency comb generation in graphene-based microresonators}

\author{Ambaresh Sahoo}
\email{ambareshs@iitg.ac.in}
\affiliation{Department of Physics, Indian Institute of Technology Guwahati, Assam 781039, India}

\begin{abstract}
Manipulation of the dynamics of cavity solitons, precise control of frequency comb spectra and nonlinear response of microresonators through exploitation of the ultrafast optical properties of graphene can have an immense impact on the technological advancement of on-chip photonic devices. Here, we report that an efficient self-frequency blueshifted frequency comb at optical and near-infrared wavelengths can be achieved owing to the near-instantaneous free-carrier dynamics of graphene in a realistic graphene-covered silicon nitride microresonator. We perform a stability analysis to find the region of stable cavity soliton excitation, which reveals that the nonlinearity of graphene helps in improving the performance of the devices.
\end{abstract}

\maketitle

\section{Introduction}
Graphene has the potential to revolutionize the speed of existing state-of-the-art devices through direct manipulation of charge carriers by ultrafast optical dynamics, which could potentially enable devices that operate at petahertz frequencies \cite{Baudisch18}. The linear and gapless electronic dispersion allows the interband optical transitions of graphene to occur at all photon energies, which makes it an excellent material exhibiting a strong nonlinear optical response \cite{Neto09}. Additionally, saturable absorption at low light intensities \cite{Marini17} and the observation of tunable localized plasmons  in doped graphene \cite{Cox17} make it superior to metal, enabling an avenue for the development of next-generation graphene-based optoelectronic and photonic devices. Ultracompact graphene-integrated waveguides \cite{Liu11,Ma20} and resonators \cite{Phare15,Wang19,Li21} are among these devices that have been proposed and analyzed with a diverse range of applications in photodetectors \cite{Mueller10}, optical filters and switches \cite{Li21}, sensors \cite{Tan21}, modulators \cite{Liu11,Phare15}, lenses \cite{Li16}, polarizers \cite{Bao11}, to name a few. 
In line with this, efforts have been put in achieving tunable fano resonances \cite{Xue20}, thermo-optic bistability \cite{Gao17}, resonant optical bistability, regenerative oscillation, and four-wave mixing \cite{Gu12}.
Although significant progress has been made in graphene-integrated hybrid microresonators and despite all-around impressive results, the ultrafast light-matter interaction between graphene and an ultra-high-Q microresonator with the full exploitation of the photo-generated graphene carriers has not been investigated, which may offer an appealing avenue for frequency comb manipulation in ultracompact optoelectronic devices.

\begin{figure}[t]
\centering
\begin{center}
\includegraphics[width=0.49\textwidth]{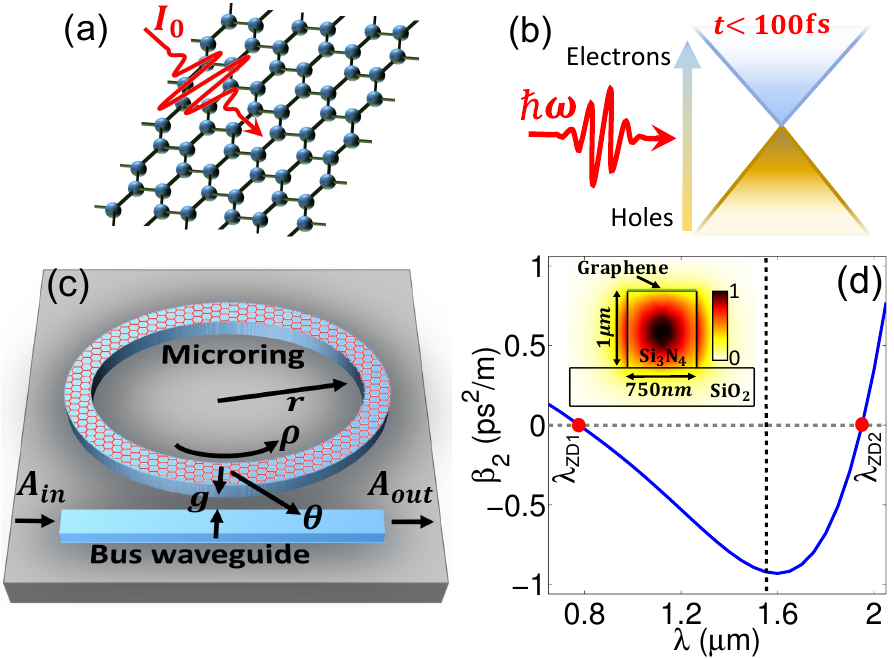}
\caption{(Color online) (a) Illustration of an extended single-layer graphene sheet illuminated by a normally-incident electric field of intensity $I_0$, and (b) instantaneous excitation of Dirac fermions in the conical band structure of graphene upon intense optical irradiation. (c) Schematic diagram of the proposed graphene-covered hybrid Si$_3$N$_4$ microresonator. (d) The cross-section of the ring with the normalized intensity of the fundamental quasi-TE mode at the input $\lambda_0=1.55\,\mu$m (vertical dashed line) and the GVD profile (solid blue curve) with the zero-dispersion wavelengths ($\lambda_{{\rm ZD}1,2}\simeq0.79,\,1.95\,\mu$m) are indicated by solid red circles.}\label{Fig1}
\end{center}\vspace{-0.6cm}
\end{figure}

In this article, we explore the dynamics of cavity solitons (CSs) and corresponding frequency combs that emerge in hybrid graphene-covered silicon nitride (Si$_3$N$_4$) microresonators by exploiting the combined effects of ultra-high-Q microresonator and ultrafast optical properties of graphene. 
The near-instantaneous free-carrier (FC)-dynamics in graphene [see Fig.\,\ref{Fig1}(a,b)] can lead to the generation of self-frequency blueshifted frequency combs at optical and near-infrared frequencies without degrading the device performances (thanks to the saturable absorption of graphene \cite{Marini17}), which cannot be directly achieved in semiconductor microresonators (due to two-photon absorption) \cite{Hansson14,Lau15,Hamerly18}. Moreover, the massless-Dirac-fermion model \cite{Baudisch18} suggests that, upon ultrafast optical excitation, a single layer of graphene is capable enough of producing $\sim 10^{12}$ carriers per cm$^2$ leading to an intense and efficient blueshift of pulse spectrum in graphene-covered hybrid waveguides than in semiconductor-based systems \cite{Sahoo21}. Also, the continuous wave (cw)-bistability analysis enables a wide range of parameter space for the CS-excitation due to the graphene carriers, which is restricted otherwise.

\section{Hybrid microresonator and the mean-field model} 
Here we propose a realistic design of hybrid graphene-covered microresonator as shown in Fig.\,\ref{Fig1}(c). The microresonator is composed of Si$_3$N$_4$ as a core material situated on SiO$_2$ with a single layer of graphene deposited on top of the Si$_3$N$_4$.  Exceptionally strong coupling of graphene carriers to optical fields produces huge number of FCs that accumulate over the resonator. From the fabrication point of view, there is an advantage of 2D material stacking on the dielectric that does not require lattice matching, which can be achieved with existing state-of-the-art facilities.  
The Si$_3$N$_4$ core is Raman inactive \cite{Guo18} and the linear and nonlinear losses are sufficiently small over the microresonator length, which is crucial for taking full advantage of the photo-generated carriers in the graphene layer to ensure optimal spectral modulation. Moreover, the intensity-dependent saturable absorption \cite{Marini17} and nonlinearity \cite{Marini16} of graphene help in manipulating the pulse dynamics and achieving low-loss propagation of optical pulses. In Fig.\,\ref{Fig1}(d), we illustrate the cross-sectional geometry with quasi-transverse-electric (TE) mode field distribution at operating wavelength $\lambda_0=1.55\,\mu$m and the group-velocity dispersion (GVD) profile of the mode over a wavelength range of $0.65-2.05\,\mu$m. At the operating wavelength (or frequency $\omega_0=2\pi c/\lambda_0$), $\beta_2\simeq -0.9852$\,ps$^2$/m, third-order Kerr nonlinear coefficient of Si$_3$N$_4$ core $\gamma\simeq1.0066$\,W$^{-1}$m$^{-1}$, and the scaling modal power $P=\int_{\rm Full~area}dx\,dy\,{\rm Re}[{\bf e}\times{\bf h}^*]\cdot\hat{\bf z}\simeq 1.8457\times10^{-12}$\,W with ${\bf e}(x,y)$ and ${\bf h}(x,y)$ being the electric and magnetic field profiles of the quasi-TE mode.

The nonlinear passive cavity dynamics in such a hybrid graphene-based microresonator [shown in Fig.\,\ref{Fig1}(c)] can be modeled through a mean-field dimensionless Lugiato-Lefever equation (LLE) (a damped-driven nonlinear Schr\"{o}dinger equation) \cite{Lugiato87,Sahoo21} which includes higher-order dispersion (HOD), Kerr and self-steepening (SS) of the Si$_3$N$_4$ core, FC-effects of graphene, saturable absorption and Kerr nonlinearity of graphene as 
\begin{align} \label{LL}
\frac{\partial u}{\partial t }=  &\left[-1 - i\,\Delta + i\sum_{n\geq2}{\delta_n \left(i \frac{\partial}{\partial\tau} \right)^n}  \right]u +S \nonumber \\
&+\,i \left[\left(1+i\tau_{sh}\frac{\partial}{\partial\tau}\right)|u|^2 u -d_{\rm FC}\Phi_{\rm FC}^c u \right.  \\ &\left. \hspace{0cm} +i\frac{\alpha_{\rm FC}^c u}{\sqrt{1+3|u|^2/u_{\rm sat}^2}}+ \frac{\alpha_{\rm FC}^c u \left(1-e^{-\eta_1\sqrt{3|u|^2/u_{\rm sat}^2}}\right)}{\sqrt{1+\eta_2\left(3|u|^2/u_{\rm sat}^2 \right)^{0.4}}}\right], \nonumber
\end{align}
where the FC-effects of graphene are included through the rate equation for the normalized carrier density $\Phi_{\rm FC}^c$ \cite{Sahoo21}
\begin{align} \label{FC_norm}
\frac{\partial {{\Phi}^c_{\rm FC}}}{\partial\tau} = \frac{\Theta_{\rm FC}^c|u|^2}{\sqrt{1+3|u|^2/u_{\rm sat}^2}}-\frac{\Phi_{\rm FC}^c}{\tau_{\rm FC}}.
\end{align} 
The rescaled parameters are followed from ref.\,\cite{Grelu_book}: Slow time $\alpha t/t_R \rightarrow t$, fast time $\tau /\tau_s\rightarrow \tau$ with $\tau_s=\sqrt{|\beta_2|L/2\alpha}$, intracavity field amplitude $u= A\sqrt{\gamma L/\alpha}$, driving field strength $S = A_{in}\sqrt{\gamma L \theta/\alpha^3}$, phase detuning $\Delta = \delta_0/\alpha$, n$^{th}$-order dispersion parameter $\delta_n=2\beta_n/\left(n!|\beta_2|\tau_s^{n-2}\right)$, SS parameter $\tau_{sh}=1/(\omega_0 \tau_s)$. Here, $t_R$, $L$, and $\alpha =(\alpha_l L+\theta)/2$ are the cavity round-trip time, cavity round-trip length,  and total cavity loss, respectively, with $\theta=0.1$ for a 90/10 coupler and $\alpha_l$ being a very small linear loss of the resonator core (Si$_3$N$_4$). The parameters related to graphene are defined in detail in ref.\,\cite{Sahoo21} and are rescaled as: FC-density $\Phi_{\rm FC}^c=K_{\rm FC} L/\alpha$, linear absorption rate $\alpha_{\rm FC}^c=w|e_0|^2 e^2 L/4\hbar P\alpha$ with $-e$ being the electron charge, saturable intensity $u^2_{sat}=3|E_{sat}/e_0|^2P\gamma L/\alpha$, FC-generation rate $\Theta_{\rm FC}^c=(\tau_s/\gamma) |e_0|^2 \pi^2{\rm v}_{\rm F} e^2/(2\hbar^2P\omega_0^2)$ with ${\rm v}_{\rm F}\simeq c/300$ being the Fermi velocity, and FC-recombination rate $\tau_{\rm FC}=1/(\gamma_{\rm FC} \tau_s)$. The realistic physical parameters, and the rescaled dimensionless parameters at the input are: Ring radius 50\,$\mu$m; $\alpha=0.2$; average $x$ component of the electric field amplitude at the middle point of the graphene--Si$_3$N$_4$ interface experienced by graphene layer $e_0\simeq (4.1731\times10^{-3} -i\,10.3177)$ V/m; $\eta_1=0.038$ and $\eta_2=0.0731$; $\alpha_{\rm FC}^c=4.135$; $d_{\rm FC}=e^2{\rm v}_{\rm F} w |e_0|^2/\pi^2\hbar\omega_0 P= 8.7729\times 10^{-7}$; $\Theta_{\rm FC}^c=1.324\times 10^7$; $u_{\rm sat}^2=0.002$; FC-recombination rate $\gamma_{\rm FC}=(100 \,{\rm fs})^{-1}$.

It is important to note that,  in our case, the free carriers in graphene are massless Dirac fermions with an effective recombination time (1/$\gamma_{\rm FC}$) $\sim100$\,fs \cite{Baudisch18}, which is significantly faster than the typical roundtrip time ($\sim 10$\,ps) of light inside a microresonator. As a result, the cumulative or roundtrip average effect of FC-accumulation over multiple roundtrips considered in the case of a silicon microresonator (because the carrier recombination time in silicon is $\sim$\,ns) \cite{Lau15,Hansson14,Hamerly18} can be ignored in the hybrid-graphene-based microresonator systems. This will lead to a carrier dynamics analogous to the linear pulse propagation in silicon waveguides  rather than circular roundtrip in silicon resonators.

\begin{figure}[t]
\centering
\begin{center}   
\includegraphics[width=0.49\textwidth]{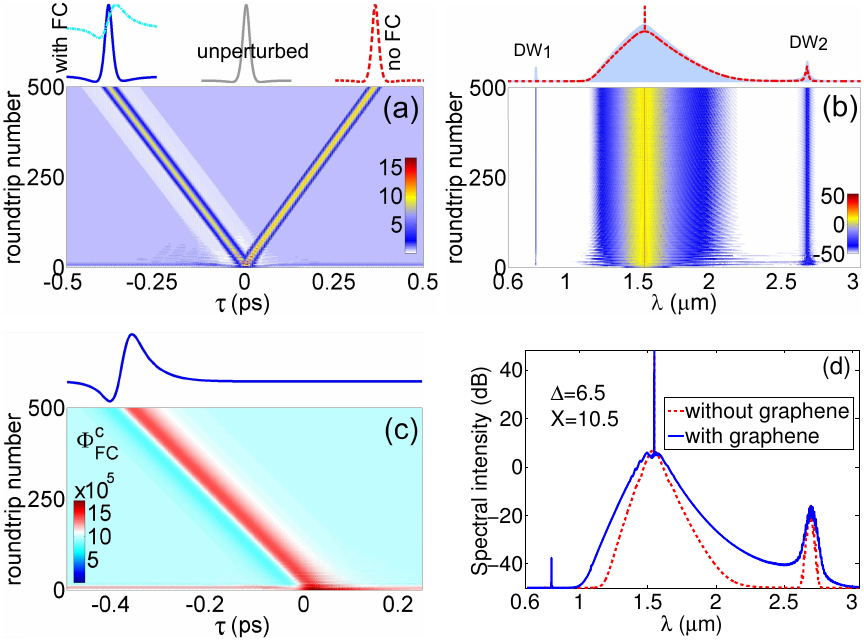}
\caption{(Color online) (a) Temporal ($|u(t,\tau)|^2$) and (b) spectral ($|\widetilde{u}(t,\omega)|^2$) evolution of CSs, and (c) evolution of $\Phi_{\rm FC}^c(t,\tau)$ over the roundtrips for $\Delta=4.5$ and $X=6.5$ with all other parameters defined above in the main text. The output profiles for different cases are shown at the top of each figure. (d) Frequency comb spectra for a different set of $X$ and $\Delta$. } \label{Fig2}
\end{center} \vspace{-0.4cm}
\end{figure}

\section{Cavity soliton and frequency comb dynamics} 
In order to investigate how the graphene carriers influence the dynamics of CSs and frequency combs, we solve the coupled LLE [Eqs.\,(\ref{LL}) and (\ref{FC_norm})] with a split-step fast Fourier technique complemented with fourth-order Runge-Kutta algorithm. Numerical solutions reveal that, due to the strong coupling with the electric field, a single layer of graphene is capable of producing enough free carriers (steady-state FC-density $\approx 10^7/cm$) that produce huge temporal and spectral modulations. 
In Fig.\,\ref{Fig2}(a) we plot the temporal evolution of CSs in the presence (negative $\tau$ axis) and in absence (positive $\tau$ axis) of graphene layer. In the Si$_3$N$_4$ resonator without graphene, we observe the usual temporal deceleration due to the HOD and SS effects (toward positive $\tau$ axis). Now, with the incorporation of graphene, the changes are dramatic, and the free carriers not only counterbalance the temporal deceleration arising from the HOD and SS but also produce huge temporal acceleration. This acceleration of the CS can be tuned further either by changing the width of the graphene layer or by changing $\tau_s$ through waveguide parameters or by changing the detuning and external pump.
At the top head of Fig.\,\ref{Fig2}(a), we depict the output profile of the temporal CSs for three cases: unperturbed (light-gray profile in the middle), where only GVD and self-phase modulation are present; without graphene (dashed red profile at the left), where HOD of Si$_3$N$_4$ and SS are present; with graphene (solid blue profile at the right). Also, we depict the normalized generated FC-density at the top left (light dot-dashed cyan curve) that influences pulse dynamics and produces this acceleration. In Fig.\,\ref{Fig2}(b), we plot the evolution of frequency comb spectrum with the influence of graphene free carriers.  The output spectrum (light blue area on the top head) demonstrates that the graphene carriers produce a significant amount of spectral modulation that generates blueshifted broadband frequency combs. As a result, the generation of dispersive wave (DW) [denoted by DW$_1$ in Fig.\,\ref{Fig2}(b) across $\lambda_{{\rm ZD}1}$] at $\lambda\simeq0.75\,\mu$m is evident, which is not possible by the Si$_3$N$_4$ resonator alone with HOD only (red dashed curve at the top). Moreover, the incorporation of graphene increases the bandwidth of the spectrum.
The temporal evolution of the normalized FC-density is also depicted in Fig.\,\ref{Fig2}(c), with the output profile on top. The free carriers produce an asymmetry in the local refractive index change \cite{Sahoo21}, pushing the CS to accelerate. As a result of this temporal acceleration in the anomalous dispersion domain, self-frequency blueshifted frequency combs are observed. Furthermore, by increasing the system parameters as shown in Fig.\,\ref{Fig2}(d), a better and efficient blueshifted comb can be obtained that reaches below $0.7\,\mu$m wavelength. All of this is possible at the near-infrared and visible wavelengths because of the saturable loss of graphene (with very low saturable intensity I$_{\rm sat}\approx 3$\,MW/cm$^2$) \cite{Marini17}, which is not possible in any semiconductor resonator due to multiphoton absorption \cite{Lau15,Sahoo19}.

\begin{figure}[t]
\centering
\begin{center}   
\includegraphics[width=0.49\textwidth]{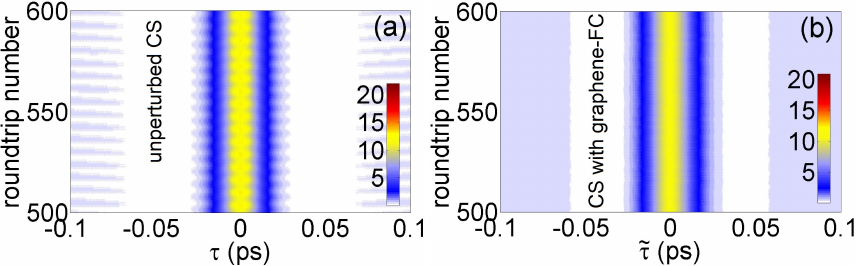}
\caption{(Color online) Evolution of (a) unperturbed CS and (b) CS with graphene carriers in a frame $\widetilde{\tau}=\tau+(\Delta \tau/\Delta t)t$ for $\Delta=5$ and $X=8$. Unperturbed CS shows periodic breathing.} \label{Fig3}
\end{center} \vspace{-0.4cm}
\end{figure}

Next, to compare what would happen if graphene carriers were the only effect, we plot the temporal evolutions while ignoring the HOD and SS terms (as is also the case with ultra-flattened GVD operating in  relatively longer pulse duration) in Figs.\,\ref{Fig3}(a) (unperturbed CS) and \ref{Fig3}(b) (CS with graphene terms in accelerated frame). Here, we observe a more robust evolution of CS in the presence of graphene than the unperturbed counterpart that shows periodic breathing. Also, a slightly lower peak intensity of the CS is evident in Fig.\,\ref{Fig3}(b) due to the absorption loss of the graphene layer. Moreover, due to the terms accounting for the graphene effects, stable CSs can be excited over a wide range of system parameters ($\Delta, X$) than the CS with typical perturbations in silicon microresonators \cite{Lau15, Sahoo19}. In the following, we will investigate these phenomena using bistability analysis.

\section{Steady-state solutions, bistability analysis, and stability of cavity solitons} 
\begin{figure}[t]
\centering
\begin{center}   
\includegraphics[width=0.49\textwidth]{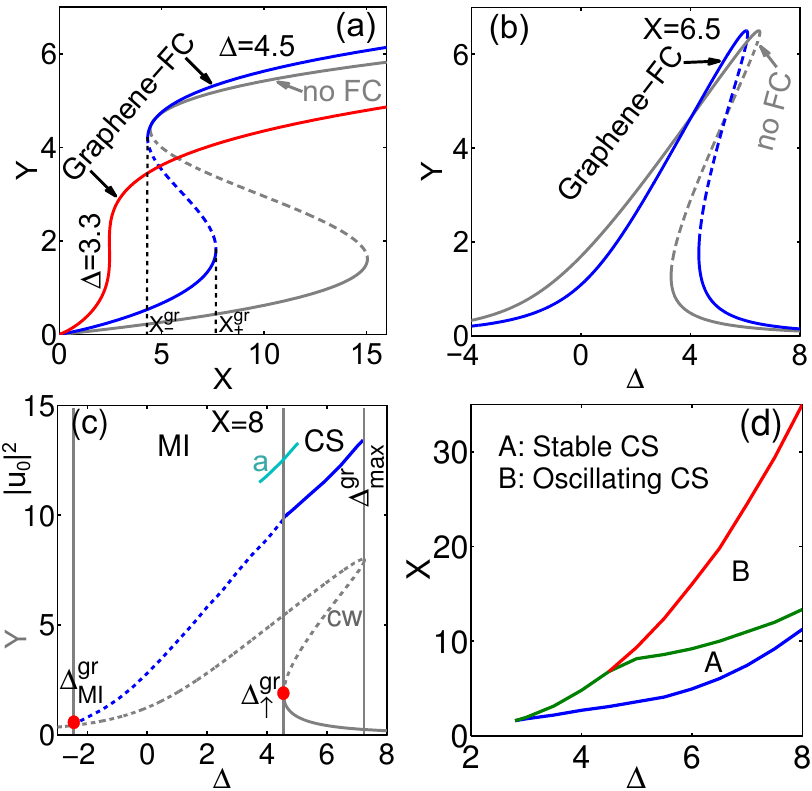}
\caption{(Color online) cw bistability curve in (a) ($X,Y$) and (b) ($\Delta,Y$) parameter space. (c) Intracavity cw solutions with MI unstable regions (gray dotted curve), the peak intensity of the MI branch (dashed blue portion), and CS branch (solid blue portion) as a function of $\Delta$ with graphene terms only. Also, the intensity of a stable CS is depicted in the absence of the graphene nonlinearity term (cyan curve `a'). (d) Attractor chart of the LLE with graphene terms, idntifying the stability region of CS.} \label{Fig4}
\end{center} \vspace{-0.4cm}
\end{figure}

To investigate deeper into the dynamics of the intracavity field, we perform a cw-bistability analysis \citep{Grelu_book}, which allows us to extract information about the parametric region of CS-excitation and frequency comb dynamics. In passive microresonators, the analytical process entails analysing the steady-state ($\partial u/\partial t=0$) and homogeneous ($\partial u/\partial \tau=0$) solution $u_s$ of the LLE [Eq.\,(\ref{LL})], which satisfies the following equation:
\begin{align} \label{OB}
X=Y\left\{1+\left[\Delta-Y+ d_{\rm FC}\left(\Phi_{\rm FC}^c\right)_{\rm s} - \mathcal{L_G} -\mathcal{K_G} \right]^2 \right\},
\end{align}
where $\mathcal{L_G}=\frac{\alpha_{\rm FC}^c }{\sqrt{1+3Y/Y_{\rm sat}}}$, $\mathcal{K_G}=\frac{\alpha_{\rm FC}^c  \left(1-e^{-\eta_1\sqrt{3Y/Y_{\rm sat}}}\right)}{\sqrt{1+\eta_2\left(3Y/Y_{\rm sat} \right)^{0.4}}}$, and the steady-state FC-density $\left(\Phi_{\rm FC}^c\right)_{\rm s}$ is evaluated from Eq.\,(\ref{FC_norm}) by setting $\partial/\partial\tau \rightarrow 0$, which takes the following form:
\begin{align} \label{FCD_steady}
\left(\Phi_{\rm FC}^c\right)_{\rm s}= \Theta_{\rm FC}^c\,\tau_{\rm FC}\,Y/\sqrt{1+3Y/Y_{\rm sat}}.
\end{align}
Here, $Y_{\rm sat}=u_{\rm sat}^2$, $Y=|u_s|^2$ and $X=|S|^2$ are the normalized intracavity and driving field powers, respectively. The bistability equation [Eq.\,(\ref{OB})] can alternatively be represented by $\Delta$ as: 
$\Delta= \left\{Y - d_{\rm FC}\left(\Phi_{\rm FC}^c\right)_{\rm s}+\mathcal{L_G}+\mathcal{K_G} \right\} \pm \sqrt{X/Y -1}$. Note that, when terms related to graphene are absent, Eq.\,(\ref{OB}) reduces to the well-known cubic bistability equation for the unperturbed LLE \cite{Grelu_book}. 
In Fig.\,\ref{Fig4}(a) and \ref{Fig4}(b), we plot the steady-state cw response of the intracavity field for the parameter spaces ($X,\,Y$) and $(\Delta,\,Y)$ using Eq.\,(\ref{OB}) and Eq.\,(\ref{FCD_steady}) by identifying the turning points $X_\pm^{\rm gr}$ that separate the upper and lower stable branches (solid portion) from the cw-unstable intermediate branch (dashed portion) which does not support any solution.  
Here, the bistability $S$ curve [Fig.\,\ref{Fig4}(a)] as well as the Kerr tilt [Fig.\,\ref{Fig4}(b)] are greatly influenced by the Kerr nonlinearity and free carriers of graphene. Also, the threshold detuning that triggers the bistability is significantly increased in the presence of graphene carriers ($\Delta_c^{\rm gr}\approx3.3$)  [solid red curve in Fig.\,\ref{Fig4}(a)] than the unperturbed case ($\Delta_c=\sqrt{3}$) \cite{Grelu_book}.
The intracavity modulation-instability analysis \cite{Grelu_book,Hansson13}, however, reveals that only the upper branch $\Delta > \Delta_\uparrow^{\rm gr}$ is stable (solid portion of the light-gray curve) and rests are unstable (dashed portion of the light-gray curve), as shown in Fig.\,\ref{Fig4}(c). In the same figure, we plot the peak intensity of the intracavity field $|u_0|^2$ as a function of detuning $\Delta$ (dotted and solid blue curve) and identify the MI region that is transitioned from the unstable lower cw branch by a critical detuning of $\Delta_{\rm MI}^{\rm gr}\approx-2.5$, the onset detuning $\Delta_{\uparrow}^{\rm gr}\approx4.72$ that stirs up the CS from MI, and the maximum detuning $\Delta_{\rm max}^{\rm gr}\approx7.172$ up to which the CS can be excited.
This plot brings us with many important remarks than in usual cases, where the perturbations degrade the duration and bandwidth of CSs \cite{Wang18} as well as the perturbations hugely affect the stability of the CS and significantly reduce the detuning range for the CS-excitation \cite{Sahoo19}. Here, although the absorption loss is present, which is saturated by the high light-field intensity (I $\gg$ I$_{\rm sat}$) because of partial Pauli blocking and makes the medium almost lossless, the nonlinearity of graphene significantly contributes to the dynamics of the CS, and as a result, it assists in improving the stability of the CS and  broadens the detuning range for the excitation of the CS. This can be easily verified by solving Eq.\,(\ref{LL}) with the graphene nonlinearity term set to zero, which results in a small range of detuning for the stable CS excitation, as shown by cyan curve `a' in Fig.\,\ref{Fig4}(c).
Next, for a complete picture of how graphene carriers affect the stability of the CS, we plot an attractor chart of the LLE in ($\Delta,X$) \cite{Leo13} revealing stability region of the CS excitation. Unlike the FC-effects of silicon, where ($\Delta,X$) is confined to a very limited region for the stable CS excitation, here, graphene carriers show a much bigger region of stable CS excitation.
Therefore, with graphene-integrated hybrid resonator, the stability of the CS improves over a wide range of detuning (which is also justified from Fig.\,\ref{Fig3}) as well as it efficiently tailors the spectrum to have a blueshifted comb at the near-infrared and visible wavelengths.

It is to be mentioned that, considering the saturable loss and nonlinearity of graphene, it is difficult to perform the MI analysis and to get exact analytical expressions of all the critical parameters, such as $\Delta_c^{\rm gr}$, $X_\pm^{\rm gr}$, $\Delta_{\rm MI}^{\rm gr}$, $\Delta_{\uparrow}^{\rm gr}$, and $\Delta_{\rm max}^{\rm gr}$ for a particular set of system parameters. 
Even in the case of silicon microresonators, the calculation involves rigorous mathematics using Lagrange's variational method with suitable approximations \cite{Sahoo19}. For this particular work, it is beyond the scope of this paper, and we leave this for future investigations.  

\section{Conclusions} 
To conclude, by exploiting the ultrafast carrier dynamics of graphene in a hybrid-graphene-based Si$_3$N$_4$ microresonator, we achieve an efficient self-frequency blueshifted frequency comb at the near-infrared and visible wavelengths with a huge temporal acceleration of the cavity soliton that can be easily manipulated. The stability analysis reveals that the nonlinearity of graphene improves the stability of the cavity soliton and thus the performance of the device. The interaction of ultrafast Dirac-fermion responses in graphene with microresonators provides an appealing optoelectronic platform with potential applications in biomedical sciences, high-sensitivity hybrid-resonator-based sensors and gyroscopes, and graphene-based lasers.

\section*{ACKNOWLEDGMENT}
A.S. acknowledges Samudra Roy (IIT-Kgp, India), Andrea Marini (University of L'Aquila, Italy), and Pascal Del'Haye (MPL, Germany) for valuable discussions.

\end{document}